\documentclass[preprint,12pt]{elsarticle}
\def\mN{\mathbb N}
\def\mZ{\mathbb Z}
\def\mR{\mathbb R}

\usepackage{amssymb}

\newtheorem{thm}{Theorem}[section]
\newtheorem{defin}[thm]{Definition}
\newtheorem{ex}[thm]{Example}
\newtheorem{rem}[thm]{Note}
 \newtheorem{lem}[thm]{Lemma}
\newcommand{\bd}{\begin{defin}}
\newcommand{\ed}{\end{defin}}
 \newcommand{\bex}{\begin{ex} \rm}
 \newcommand{\eex}{\end{ex}}
 \newcommand{\brem}{\begin{rem} \rm}
 \newcommand{\erem}{\end{rem}}
  \newcommand{\blem}{\begin{lem}}
 \newcommand{\elem}{\end{lem}}
  \newcommand{\pr}{\noindent{\bf Proof. }}
 \newcommand{\ep}{\nolinebreak{\hspace*{\fill}$\Box$ \vspace*{0.25cm}}}

\journal{Computational Statistics \& Data Analysis}

\begin{document}

\begin{frontmatter}



\title{Nonsensical Excel and Statistica Default Output and Algorithm for an Adequate Display}


\author[smk]{Sne\v zana Mati\' c-Keki\' c}
\author[nd]{Neboj\v sa Dedovi\'c}
\author[bm]{Beba Mutavd\v zi\'c}
\address[smk]{University of Novi Sad, Faculty of Agriculture, Department for Agricultural Engineering, Tel.: +381-21-4853283, e-mail: snezana.kekic@polj.uns.ac.rs, Trg Dositeja Obradovi\'ca 8, 21000 Novi Sad, Serbia}
\address[nd]{University of Novi Sad, Faculty of Agriculture, Department for Agricultural Engineering, Tel.: +381-21-4853292, e-mail: dedovicn@polj.uns.ac.rs, Trg Dositeja Obradovi\'ca 8, 21000 Novi Sad, Serbia}
\address[bm]{University of Novi Sad, Faculty of Agriculture, Department for Economy, Tel.: +381-21-4853382, e-mail: bebam@polj.uns.ac.rs, Trg Dositeja Obradovi\'ca 8, 21000 Novi Sad, Serbia}

\begin{abstract}
The purpose of this paper is to present an algorithm that determines the necessary and sufficient number of significant digits in the coefficients of polynomial trend in order to achieve a pre-specified precision of polynomial trend. Thus, the obtained coefficients should be presented in the default output when fitting the experimental data by polynomial trend. Namely, in order to find the best fitting function for certain type of data there is a real possibility of making significant errors using default output of Excel 2003, 2007, 2010, 2013 and Statistica for Windows 2007-2012 software packages. Conversely, software package Mathematica (version 6) has shown very good characteristics in dealing with precision problems, although the default output sometimes shows more significant digits than necessary. Also, it turned out that the software packages Excel and Statistica violated the order of operations as defined in mathematics. For example, $-1^2=1$ and $-(-3)^2=9$. This problem always occurs in Excel, while at Statistica during the graphical interpretation of a function.
\end{abstract}


\end{frontmatter}


\section{Introduction}
Sometimes, we are faced with different mathematical problems which are partially or completely unsolvable without using a PC. For example, it helps us to generate large number (approximately 2000) of new 2-designs \cite{ack00} or completely solve the problem of optimal construction of digital convex polygons with $n$ vertices, $n\in \mN$ \cite{mat96,mat97}. Also, it is well known that the analysis of experimental data must be done by using the software support utilities. This is supported by the fact that tools for data processing have been developed enormously for the last twenty years. However, due to financial, hardware or other limitations, we tend to use some preferred software packages that we are familiar with, instead of applying better software tools for application in our field of research. 

The problems of experimental data analysis with adequate software were mentioned by several authors (\cite{harg10,hes06,stok04,spt11}). Hargreaves and McWilliams (\cite{harg10}) pointed out the problems with implementation of Excel 2007 software package when generating polynomial trend line equations. According to these authors, Excel will "fit" nonsensical trend lines to data presented in column and line charts, and it can report an inadequate number of significant digits for polynomial trend lines. The authors also offer possible solutions to fix these problems. Hesse (\cite{hes06}) indicated that the analysis of experimental data may give incorrect linear trend line and an answer to the question why the use of logarithmic transformation of data is preferable. Stokes (\cite{stok04}) suggested that the problem of data analysis should be solved with two or more software packages because some of the used software gave different results. The author suggested the need for further analysis of such problems because the software might contain subtle numerical problems that were not always obvious.

Basically, this study includes data analysis by the software packages Mathematica 6 \cite{wolf03}, Microsoft Office Excel 2003, 2007, 2010, 2013 \cite{exe03} and Statistica for Windows 2007-2012 \cite{sta10}. Referring to the fruit drying process \cite{mosh80}, it was necessary to calculate the surface area of a pear, i.e. integral of a non-negative fitting function obtained by the Excel software, but the result was nonsensical-the negative number. Similarly, odd results was obtained using the fitting function in the software package Statistica. It is obtained three times larger volume of a pear than the actual one obtained by Archimedes method. The problem was the fact that the coefficients of the fitting functions had an insufficient number of significant digits, which is demonstrated in this paper. Hargreaves and McWilliams (\cite{harg10}) noticed similar in Excel software package. They suggested that the number of significant digits of the coefficients of polynomial trends should be increased, and that this number should be the same for all coefficients. On the other hand, Mathematica showed good default output characteristics, especially from the aspect of enough number of significant digits, but sometimes the default output shows more significant digits than necessary.

This paper presents an algorithm that provides a necessary and sufficient number of significant digits, which is generally not the same for all coefficients of the polynomial trend. The algorithm was tested on three data sets. Additionally, this study contains a Note which includes data refers to the the straw bale combustion process \cite{brk06,ded12}. After fitting these experimental data obtained in the combustion of wheat straw bales \cite{ded12}, the resultant fitting function from Statistica did not match either experimental data or coefficient of determination. Although the regression coefficients were calculated correctly and tested in Mathematica, the graphic of the fitting function was incorrect. Since we could not find the cause of this problem in the first steps, we decided to fit the experimental data by Excel, as a form of simpler and widely used software application. This helped us understand better the causes of this problem, as it will be described later. It turned out that the software packages Excel and Statistica violated the order of operations as defined in mathematics. For example, $-x^2$ must be $-1\cdot x^2$, but in Excel and Statistica program packages it is $(-x)^2$.

\section{Results and discussion}
For the sake of algorithm simplicity and its further application, without loss of generality, it can be assumed that the experimental data are given as
\begin{equation}\label{usl}
0\leq x_0<x_1<\dots<x_n,\; y_i\geq 0 \mbox{ za } i=0,1,\dots,n.
\end{equation}

\begin{figure}[htbp]
  \includegraphics[width=0.9\textwidth]{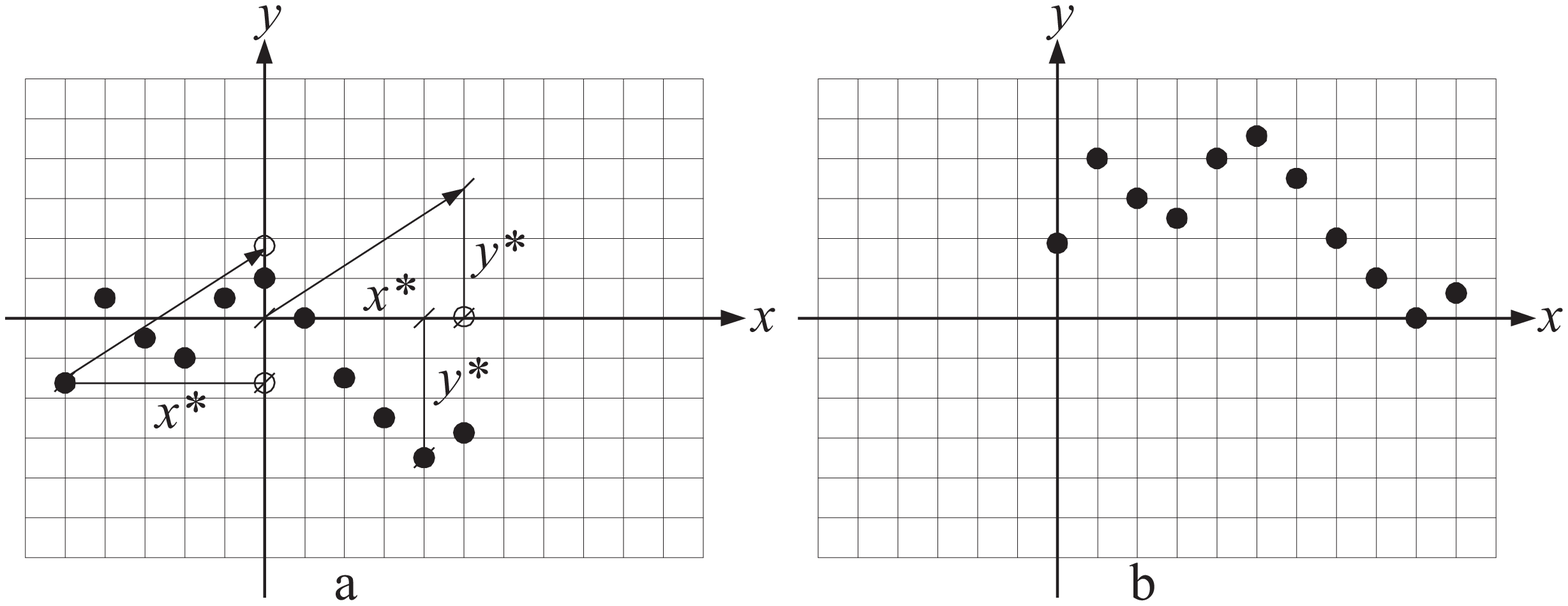}
\caption{$(x^*_i,y^*_i)$ (a, solid symbols) translation by vector $(x^*,y^*)$ into non-negative $(x_i,y_i)$ (b, solid symbols), where $x^*=\max_{x^*_i<0}|x^*_i|$ and $y^*=\max_{y^*_j<0}|y^*_j|$, $i=0,1,\dots n$.}
\label{trans}       
\end{figure}

 Namely, let the experimental data $(x^*_i,y^*_i)$, $i=0,1,\dots n$ (Figure \ref{trans}a) which do not satisfy (\ref{usl}) be given. Hence, let $x^*_i$ exist for $i\in I\subseteq\{0,1,\dots n\}$ such that $x^*_i<0$. For those $i\in I$, we determine $\displaystyle x^*=\max_{i\in I}|x^*_i|$. Analogously, let $y^*_j$ exist for $j\in J\subseteq\{0,1,\dots n\}$ such that $y^*_j<0$. Again, for those $j\in J$, we determine $\displaystyle y^*=\max_{j\in J}|y^*_j|$. After $(x^*_i,y^*_i)$ translation for vector $(x^*,y^*)$, we obtain the points $(x^*_i+x^*,y^*_i+y^*)=:(x_i,y_i)$ where $x_i,y_i\geq 0$ holds for $i=0,1,\dots n$ (Figure \ref{trans}b). It can be concluded that arbitrary experimental data $(x^*_i,y^*_i)$ can always be translated to $(x_i,y_i)$, $i=0,1,\dots n$, such that $x_i\geq 0$ and $y_i\geq 0$ holds for $i=0,1,\dots n$. If $x_i$ are not sorted, then sorting $(x_i,y_i)$ by the first coordinate, one can provide that (\ref{usl}) holds. In this paper, all experimental data satisfy (\ref{usl}).

%

Now, it will be given an example where the problem with poor default output of polynomial trend was noticed.

\begin{figure}[htbp]
  \includegraphics[width=0.9\textwidth]{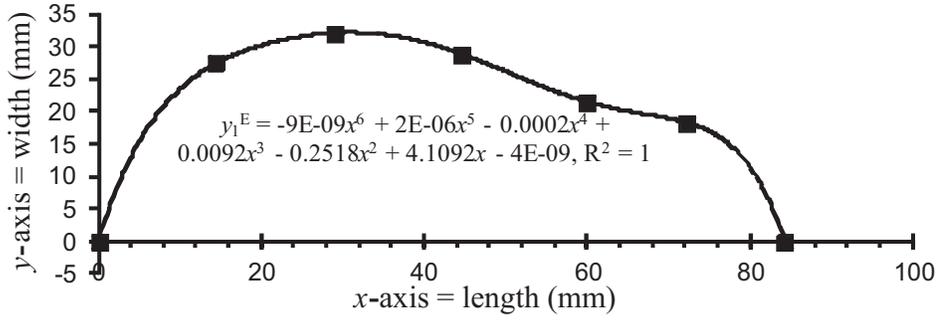}
\caption{The correct graph for pear border line which passes through the average experimental points $T_i$, $i=0,1,\dots,6$ (black squares) in Excel, with incorrect reported equation. Presented fitting function for pear border line, $y_1^E(x)$, does not suit reported graphic.}
\label{1a}       
\end{figure}

\bex\label{krus}
In \cite{bab11,ded11}, it was necessary to identify the function which would approximate border line of an average Williams pear as precisely as possible,
because it was further used for the calculation of pear surface area and volume. Consequently, 6th order polynomial function $y_1(x)=\sum_{k=0}^{6} {a_{k}}\,x^{k}$ was chosen. It passed through the points of the outer edge of analyzed pear contour $T_0(0, 0)$, $T_1(14.5, 27.4)$, $T_2(29, 32.1)$, $T_3(44.5, 28.9)$, $T_4(60, 21.5)$, $T_5(72.2, 18.3)$ and $T_6(84.3, 0)$. These points are marked as solid symbols (Figure \ref{1a}).
%

 The polynomial fitting function for data $T_i(x_i,y_i)$, $i=0,1,\dots,6$ was determined separately in program packages Mathematica, Statistica and Excel for Windows and was marked as $y_1^M(x)$, $y_1^S(x)$ and $y_1^E(x)$-ver. 2003, 2010, 2013 ($y_1^{E^*}(x)$-ver. 2007), respectively. Coefficients of these polynomials are given in Table \ref{t2}.

 \begin{table}[htbp]
\caption{Values of the coefficients $a_{k}$ (default output), $k=0,1,\dots,6$, of polynomial $y_{1}^{W}(x)=\sum_{k=0}^{6} a_{k}\,
x^{k}$ obtained from polynomial interpolation in Mathematica ($W=M$) and from fitting function in Excel ($W=E$ ver. 2003,2010,2013, $W=E^{*}$ ver. 2007) and regression function in Statistica ($W=S$, ver. 2007-2012) for data of pear border $T_i(x_i,y_i)$. $y^A_1$ - polynomial obtained after applying proposed algorithm with optimal coefficients.}
\label{t2}       
{\small \begin{tabular}{cccccc}
\hline\noalign{\smallskip}
 & $y_{1}^{E}$ & $y_{1}^{E^{*}}$ & $y_{1}^{S}$ & $y_{1}^{M}$ & $y^A_1$ \\
\noalign{\smallskip}\hline\noalign{\smallskip}
$a_{6}$ & $-9\cdot10^{-9}$   & $-9\cdot10^{-9}$  & $-8.5\cdot10^{-9}$   & $-8.51613\cdot10^{-9}$  & ${\bf -8.5161\cdot 10^{-9}}$  \\
$a_{5}$ & $2\cdot10^{-6}$    & $2\cdot10^{-6}$   & $2.06\cdot10^{-6}$   & $2.06269\cdot10^{-6}$   & ${\bf 2.06269\cdot 10^{-6}}$  \\
$a_{4}$ & $-0.0002$          & $-0.000$          & $-1.9\cdot 10^{-4}$  & $-1.94089\cdot10^{-4}$  & ${\bf -1.94089\cdot 10^{-4}}$ \\
$a_{3}$ & $0.0092$           & $0.009$           & $9.213\cdot 10^{-3}$ & $9.21349\cdot10^{-3}$   & ${\bf 9.2135\cdot 10^{-3}}$   \\
$a_{2}$ & $-0.2518$          & $-0.251$          & $-0.25177$           & $-0.251775$             & ${\bf -0.25178}$              \\
$a_{1}$ & $4.1092$           & $4.109$           & $4.10924$            & $4.10924$               & ${\bf 4.109}$                 \\
$a_{0}$ & $-4\cdot10^{-9}$   & $2\cdot10^{-10}$  & $-0.0\cdot10^{-21}$  & $-5.35881\cdot10^{-13}$ &                               \\
\noalign{\smallskip}\hline
\end{tabular}
}
\end{table}

\begin{figure}[htbp]
  \includegraphics[width=0.75\textwidth]{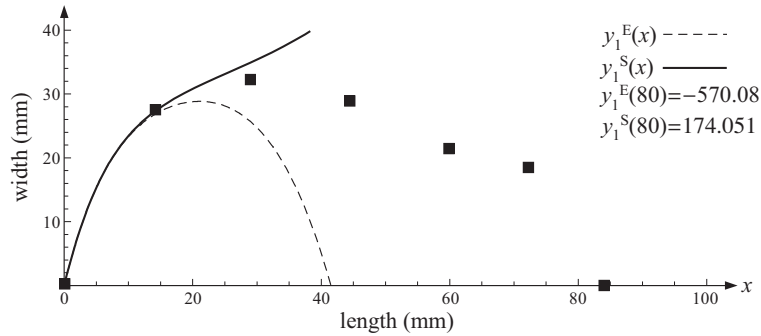}
\caption{Fit by reported equations $y_1^{E}(x)$ and $y_1^{S}(x)$ drawn in Mathematica 6.}
\label{1b}       
\end{figure}

 Unfortunately, equations of polynomial trend lines obtained by Statistica and Excel, which was reported as default output (Table \ref{t2}), do not fit their experimental data (Figure \ref{1b}). Further usage of these functions would create very odd results (for example, negative values for the surface area and volume of rotating body).
 %

Obviously, the coefficients of the following polynomial $y_1^E(x)\,(y_1^{E^*}(x))$, $y_1^{S}(x)$ and $y_1^M(x)$ which multiply the same term $x^k$, $k=1,\dots,6$, are similar (Table \ref{t2}). According to the same table, it is also evident that constant terms are different, but their order is too small and do not have large influence on the lack of precision for $y_1^E(x)\,(y_1^{E^*}(x))$ and $y_1^{S}(x)$. Clearly, that poor precision is caused by badly rounded coefficients which multiply $x^3$, $x^4$, $x^5$ and $x^6$. For example, for $x=80$, $x^6$ is value of order $10^{12}$ and the digits of $a_6$ at positions from $10^{-9}$ to $10^{-13}$ become significant, but those digits are missing from the coefficient $a_6$ in $y_1^E(x)\,(y_1^{E^*}(x))$ and $y_1^{S}(x)$ (Table \ref{t2}).
\eex

\subsection{Two ways to overcome the problems associated with default output of the polynomial trend coefficients}

The first way is to use presented algorithm which determines the number of significant digits necessary to be reported in the default output of each polynomial coefficient, while the second one, so-called the 'First Aid', is to increase the number of significant digits using the options offered in the observed software packages.

\subsubsection{The algorithm for precise determination of the number of significant digits in the coefficients of polynomial trends}

 \bd\label{d1} Each $z\in\mR^+\cup\{0\}$ can be written as
 $$
 z=\sum_{i=-\infty}^{r}z_i\cdot 10^i,\;z_i\in\{0,1,\dots,9\},\;z_r\neq 0,
 $$
 but since we are dealing with the experimental data, they can always be rounded such that if $10^P$ is the position of the last significant digit of the experimental data $z$, then
 $$
 z=\sum_{i=P}^{r}z_i\cdot 10^i,\;z_i\in\{0,1,\dots,9\},\;z_r\neq 0.
 $$
 We shall write ${\cal P}(z)=10^P$, and this is the precision of $z$. The error is then at $10^{P-1}$ position.
 \ed

Thus, the total number of significant digits of $z$ is equal to $r-P+1$. Let us give two examples. If $z=12000$ and 2 is the last significant digit, then, according to the definition (\ref{d1}) $r=4$ and $P=3$. But, if $z=0.0123$, then $r=-2$ and $P=-4$.

Now, let $y(x)=\sum_{k=0}^n a_k\,x^k$ be a polynomial fitting function of data $(x_i,y_i)$, $i=0,1,\dots n$, such that (\ref{usl}) holds. Suppose that this function is determined by mentioned software packages. Let $y^A(x)=\sum_{k=0}^{d} {\bf a_{k}}\,x^{k}$, $x\in [x_0,x_n]$, $d\leq n$ be a polynomial which coefficients ${\bf a_k}$, $k=0,1,\dots,d$ are obtained from the coefficients $a_k$, $k=0,1,\dots,d$ of polynomial $y(x)$ after applying proposed algorithm given bellow. The coefficients ${\bf a_k}$, $k=0,1,\dots,d$ will be regarded as optimal ones because they will have minimal number of precisely determined significant digits. This number will be enough to maintain the accuracy of the fitting polynomial trend (definition \ref{def1}).

 \bd\label{def1} Coefficients ${\bf a_k}$ of the polynomial fitting function $y^A(x)$ are optimal ones if
 \begin{itemize}
 \item[(i)] $\left|y_i-y^A(x_i)\right|=c\cdot 10^{P-1},\hspace{1.3cm}i=0,1,\dots,n,\;(d=n)$, or
  \item[(ii)] $\left|y_i-y^A(x_i)\right|=c\cdot 10^{P-1}\cdot 10^B,\hspace{0.3cm}i=0,1,\dots,n,\;(d<n)$,
 \end{itemize}
 where $P\in\mZ$, $1\leq c<10$, $10^B$ is the position of the first digit of $(1-R^2)\cdot y_{\max}$, $\displaystyle y_{\max}=\max_{i\in\{0,\dots,n\}}y_i$, $R^2$ is the coefficient of determination of the fitting polynomial $y(x)$, and $10^P$ is the precision of the measured data $y_i$ in the point $x_i$, for $i=0,1,\dots,n$.
 \ed

\brem\label{n1} Precision of the sum $y^A(x)=\sum_{k=0}^{d} {\bf a_{k}}\,x^{k}$, $x\in [x_0,x_n]$, is at most equal to precision of the most imprecise addend. Therefore, we require that all addends ${\bf a_{k}}\,x^k$ have the same precision equal to $10^{P-1}$ because of possible error accumulation when summarizing.
\erem

In the sequel, we will assume that all experimental data $x_i$, $i=0,1,\dots,n$, are written with correct digits only with the same precision equal to $10^{P_x}$. Of course, $10^{P_x}$ is the position of the last correct digit in this case.

\blem\label{lema1} Suppose that $0\leq x_0<x_1<\dots<x_n$. Then for $x_i\in[x_0,x_n]$, number $x_n^k$, $k\in\mN$, has the largest number of incorrect digits among numbers $x_i^k$, $i=0,1,\dots,n$.
\elem
\pr It will be shown that $x_l^k$ has at least the same number of incorrect digits as $x_{l-1}^k$ for arbitrary $l\in\{1,2,\dots,n\}$. Then, it will follow that $x_n^k$ has at least the same number of incorrect digits as $x_{l}^k$, $l=0,1,\dots,n$.

Suppose that $x_l$ can be written as
$$
x_l=\sum_{i=P_x}^{r_l}x_{l,i}\cdot 10^i,\;x_{l,i}\in\{0,1,\dots,9\},\;x_{l,r_l}\neq 0\,.
$$
Then, the number of correct digits is equal to $r_{l}-P_x+1$. Again, we assume that $x_{l-1}$ can be also written as
$$
x_{l-1}=\sum_{i=P_x}^{r_{l-1}}x_{l-1,i}\cdot 10^i,\;x_{l-1,i}\in\{0,1,\dots,9\},\;x_{l-1,r_{l-1}}\neq 0.
$$
Thus, the number of correct digits is equal to $r_{l-1}-P_x+1$. $10^{P_x}$ is the position of the last correct digit of the $x_{l-1}$ and $x_l$.

Clearly, $r_{l-1}\leq r_{l}$ and it is well known that if $x$ is written with correct digits only, then $x^k$, $k\in\mN$, has the same number of correct digits as $x$.

{\it Case 1.} Let $x_l$ and $x_{l-1}$ have the first correct digit at the same position, i.e. let $r_{l-1}=r_{l}$. Maximum number of digits of $x_{l-1}^k$ is equal to $k\cdot (r_{l-1}-P_x+1)$, while the number of correct digits of $x_{l-1}$ is equal to $(r_{l-1}-P_x+1)$. Consequently, $x_{l-1}^k$ has $(r_{l-1}-P_x+1)$ correct digits, implying that number of incorrect digits of $x_{l-1}^k$ is equal to
\begin{equation}\label{bsc}
k\cdot (r_{l-1}-P_{x}+1)-(r_{l-1}-P_{x}+1)\,.
\end{equation}
Also, $x_l>x_{l-1}$ implies that number of digits of $x_{l}^k$ is equal to $k\cdot (r_{l}-P_{x}+1)$ so the number of incorrect digits is
$$
k\cdot (r_{l}-P_{x}+1)-(r_{l}-P_{x}+1)\,.
$$
This means that $x_l^k$ and $x_{l-1}^k$ have the same number of incorrect digits since $r_{l-1}=r_{l}$.

{\it Case 2.} If $r_l\geq r_{l-1}+1$, then the maximum number of incorrect digits of $x_{l-1}^k$ is given in (\ref{bsc}), while the minimum number of incorrect digits of $x_{l}^k$ is
$$
\left(k\cdot (r_{l}-P_x+1)-(k-1)\right)-(r_{l}-P_x+1)\,.
$$
Namely, among numbers with the same number of digits, 10 to the power of integer contains minimum number of digits. For example, $10^s$ has $(s+1)$ digits, while $(10^s)^k$ contains $(s\,k+1)$ digits, i.e. $(k\cdot (s+1) - (k-1))$. So,
$$
\begin{array}{l}
\left(k\cdot (r_{l}-P_x+1)-(k-1)\right)-(r_{l}-P_x+1)\\
\hspace*{3cm}= r_{l}(k-1)-P_x(k-1)\\
\hspace*{3cm}\geq (r_{l-1}+1)(k-1)-P_x(k-1)\\
\hspace*{3cm}=k\cdot (r_{l-1}-P_x+1)-(r_{l-1}-P_x+1)\,.
\end{array}
$$
Here, we have used $r_l\geq r_{l-1}+1$. Conclusion is that the number of incorrect digits of $x_l^k$ is greater or equal to the number of incorrect digits of $x_{l-1}^k$. Thus, lemma is proved.
\ep

 Next, we will write
 $$
 x_n=\bar c\cdot 10^{\bar P},\;\;0.1\leq \bar c<1,\;\bar P\in\mZ.
 $$
If the precision of the measured data is $10^P$, according to Note \ref{n1}, it is necessary and sufficient that the precision of the addends of interpolation polynomial trend be $10^{P-1}$. Consequently, we require that precision of ${\bf a_{k}}\,x^k$ be $10^{P-1}$, i.e. based on Lemma \ref{lema1}, we have
 \begin{equation}\label{prec}
 {\cal P}({\bf a_{k}}\,x^k)={\cal P}({\bf a_{k}}\,x_n^k)={\cal P}({\bf a_{k}}\,(\bar c\cdot 10^{\bar P})^k)=10^{P-1}\,.
 \end{equation}

If we denote ${\bar c}^{k}$ as $10^{P_{\bar c}}$, then $P_{\bar c}=k\,\log_{10}\,\bar c$. The position of the first significant digit of $x_n^k$ is then $10^{[P_{\bar c}+k\,\bar P]}$, where $[x]$ is the nearest integer to $x$. Now, it is necessary to show that the position of the last significant digit of ${\bf a_{k}}\,x_n^k$, ${\cal P}({\bf a_{k}}\,x_n^k)$, is equal to the position of the last significant digit of ${\bf a_{k}}$, ${\cal P}({\bf a_{k}})$, multiplied by the position of the first significant digit of $x_n^k$. But it is true, because the only digit of $x_n^k$, that is always correct is the first one at position $10^{[P_{\bar c}+k\,\bar P]}$. The rest of the digits might be the incorrect ones, which multiplication with the correct digits gives the incorrect ones. Therefore, (\ref{prec}) implies
 $$
 {\cal P}({\bf a_{k}}\,x^k)={\cal P}({\bf a_{k}})\,{\bar c}^k\cdot 10^{k\,\bar P}=10^{P-1}\Rightarrow {\cal P}({\bf a_{k}})=10^{-P_{\bar c}}\cdot 10^{P-1-k\,\bar P}\,.
  $$

 From the above mentioned, the {\bf algorithm} can be written as follows: after fitting data by $y^A(x)=\sum_{k=0}^{d} {\bf a_{k}}\,x^{k}$, $x\in [x_0,x_n]$, the precision of ${\bf a_k}$ will be determined such that the last significant digit be at position
 \begin{itemize}
 \item[(i)] $10^{[P-1-k\,\bar P-P_{\bar c}]}$, $d=n$ (interpolation polynomial);
 \item[(ii)] $10^{[P-1+B-k\,\bar P-P_{\bar c}]}$, $d<n$,
 \end{itemize}
 where:
 \begin{itemize}
 \item $[x]$ is the nearest integer to $x$,
 \item $P$ is the position of the last significant digit of measured data $y_i$ in the point $x_i$, for $i=0,1,\dots,n$,
 \item $\bar P$ and $\bar c$ are obtained from $x_n=\bar c\cdot 10^{\bar P},\;\;0.1\leq \bar c<1,\;\bar P\in\mZ$,
 \item $P_{\bar c}=k\,\log_{10}\,\bar c$, $k=0,1,\dots,d$.
 \item $10^B$ is the position of the first significant digit of the $(1-R^2)\cdot y_{\max}$, $y_{\max}=\max_{i\in\{0,\dots,n\}}y_i$.
 \end{itemize}

 \subsection{Application of the algorithm on the interpolating polynomial ($d=n$)}
 In the Example \ref{krus}, $P=0$ because the experimental data were integers, with the last significant digit at the position $10^0$. Then, the average values of $y_i$ at the points $T_i(x_i,y_i)$, $i=0,1,\dots,6$ were calculated. Therefore, $x_6=84.3=0.843\cdot 10^2\Rightarrow \bar c=0.843$, $\bar P=2$. Hence, the polynomial coefficients should be reported with the following precision: ${\cal P}({\bf a_0})=10^{-1}$, ${\cal P}({\bf a_1})=10^{-3}$, ${\cal P}({\bf a_2})=10^{-5}$, ${\cal P}({\bf a_3})=10^{-7}$, ${\cal P}({\bf a_4})=10^{-9}$, ${\cal P}({\bf a_5})=10^{-11}$ and ${\cal P}({\bf a_6})=10^{-13}$ (Tables \ref{t2} and \ref{t3}).

 \begin{table}[htbp]
\caption{Precision of the coefficients of polynomial (\ref{E1}), i.e. position of the last significant digit, $x_6=84.3=0.843\cdot 10^2\Rightarrow \bar c=0.843$, $\bar P=2$, $P_{\bar c}=k\,\log_{10}\,\bar c$.}
\label{t3}       
{\small
\begin{tabular}{cccccccc}
\hline\noalign{\smallskip}
 & $k=0$ & $k=1$ & $k=2$ & $k=3$ & $k=4$ & $k=5$ & $k=6$\\
\noalign{\smallskip}\hline\noalign{\smallskip}
$P_{\bar c}$ & $0$   & $-0.074$   & $-1.148$ & $-0.223$ & $-0.297$ & $-0.371$ & $-0.445$\\
$P-1-k\,\bar P-P_{\bar c}$ & $-1$   & $-2.93$   & $-4.85$ & $-6.78$ & $-8.70$ & $-10.63$ & $-12.56$\\
$[P-1-k\,\bar P-P_{\bar c}]$ & $-1$   & $-3$   & $-5$ & $-7$ & $-9$ & $-11$ & $-13$\\
Mathematica & -   & $-5$   & $-6$ & $-8$ & $-9$ & $-11$ & $-14$\\
\noalign{\smallskip}\hline
\end{tabular}
}
\end{table}

 The polynomial trend should be displayed as
 \begin{equation} \label{E1}
 \begin{array}{rcl}
 y^A_1(x) & = & -8.5161\cdot 10^{-9} x^6+2.06269\cdot 10^{-6} x^5-1.94089\cdot 10^{-4} x^4\\
 &  & +9.2135\cdot 10^{-3} x^3 - 0.25178\,x^2+4.109\,x\,,
 \end{array}
 \end{equation}
 where the coefficients are shown with significant digits only.

\begin{figure}[htbp]
  \includegraphics[width=0.77\textwidth]{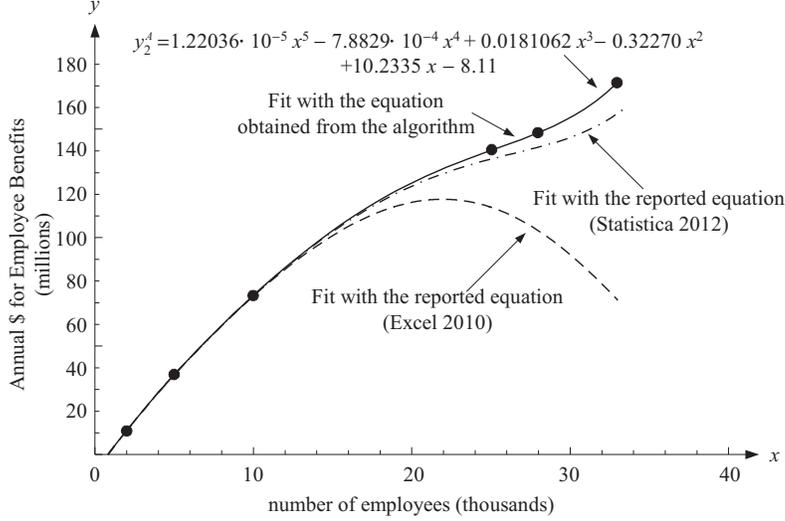}
\caption{5th order polynomial trend line equation reported by Excel 2010 ($y^E_2=1\cdot 10^{-5}x^5 - 0.0008 x^4 + 0.0181 x^3 - 0.3227 x^2 + 10.234 x - 8.1089$, dashed curve) and Statistica 2012 ($y^S_2=1.2204\cdot 10^{-5}x^5 - 0.0008 x^4 + 0.0181 x^3 - 0.3227 x^2 + 10.2335 x - 8.1089$, dash-dot curve). A polynomial trend line obtained from the algorithm (solid curve) and its equation.}
\label{empl}       
\end{figure}

\bex To confirm the validity of proposed algorithm, it will be applied on the data from Hargreaves and McWilliams (\cite{harg10}). The authors supposed that they have data on the number of employees at six companies expressed in units of 1000, and also on the annual cost of employee benefits for these companies, expressed in units of one million as follows: $(2,11.2)$, $(5,36.8)$, $(10,73.4)$, $(25,140.2)$, $(28,148.4)$, $(33,171.6)$ (Figure \ref{empl}, solid symbols). %

The fifth order polynomial trend (which corresponds to the interpolating polynomial) reported by default output of Excel 2010 is $y^E_2=1\cdot 10^{-5}x^5 - 0.0008 x^4 + 0.0181 x^3 - 0.3227 x^2 + 10.234 x - 8.1089$. Fit by the reported equation is at Figure \ref{empl} (dashed curve). The fifth order polynomial trend for the same data, but after application of the proposed algorithm (Table \ref{t4}, first part) is
 \begin{equation} \label{E2}
 \begin{array}{rcl}
y^A_2&=&1.22036\cdot 10^{-5}x^5-7.8829\cdot 10^{-4}x^4+0.0181062 x^3\\
     &&- 0.32270 x^2+ 10.2335 x-8.11
 \end{array}
 \end{equation}
because $P=-1$, $x_{\max}=33$, $\bar c=0.33$ and $\bar P=2$ (Figure \ref{empl}, solid line).

\begin{table}[htbp]
\caption{Precision of the coefficients of polynomials (\ref{E2}) and (\ref{E3}) for the same experimental data $(x_k,y_k)$, $a^u_k$ - coefficients from (\ref{E2}) ($x_k$ expressed in units of 1000, $y_k$ expressed in units of one million), $a_k$ - coefficients from (\ref{E3}) ($x_k$ and $y_k$ expressed in units of 1), $k=0,1,\dots,5$, $P_{\bar c}=k\,\log_{10}\,\bar c$.}
\label{t4}       
\begin{tabular}{ccccccc}
\hline\noalign{\smallskip}
{\scriptsize $P=-1$, $\bar c=0.33$, $\bar P=2$} & ${\cal P}(a^{u}_0)$ & ${\cal P}(a^{u}_1)$ & ${\cal P}(a^{u}_2)$ & ${\cal P}(a^{u}_3)$ & ${\cal P}(a^{u}_4)$ & ${\cal P}(a^{u}_5)$\\
\noalign{\smallskip}\hline\noalign{\smallskip}
$10^{[P-1-k\,\bar P-P_{\bar c}]}$ & ${\bf 10^{-2}}$   & ${\bf 10^{-4}}$   & ${\bf 10^{-5}}$ & ${\bf 10^{-7}}$ & ${\bf 10^{-8}}$ & ${\bf 10^{-10}}$ \\
Excel 2010 & $10^{-4}$   & $10^{-3}$   & $10^{-4}$ & $10^{-4}$ & $10^{-4}$ & $10^{-5}$ \\
Statistica 2012 & $10^{-4}$   & $10^{-4}$   & $10^{-4}$ & $10^{-4}$ & $10^{-4}$ & $10^{-9}$ \\
Mathematica & $10^{-5}$   & $10^{-4}$   & $10^{-6}$ & $10^{-7}$ & $10^{-9}$ & $10^{-10}$ \\
\noalign{\smallskip}\hline
{\scriptsize $P=5$, $\bar c=0.33$, $\bar P=5$} & ${\cal P}(a_0)$ & ${\cal P}(a_1)$ & ${\cal P}(a_2)$ & ${\cal P}(a_3)$ & ${\cal P}(a_4)$ & ${\cal P}(a_5)$\\
\noalign{\smallskip}\hline\noalign{\smallskip}
$10^{[P-1-k\,\bar P-P_{\bar c}]}$ & ${\bf 10^{4}}$   & ${\bf 10^{-1}}$   & ${\bf 10^{-5}}$ & ${\bf 10^{-10}}$ & ${\bf 10^{-14}}$ & ${\bf 10^{-19}}$ \\
Excel 2010 & $10^{6}$   & $10^{0}$   & $10^{-4}$ & $10^{-5}$ & $10^{-10}$ & $10^{-14}$ \\
Statistica 2012 & $10^{2}$   & $10^{-4}$   & $10^{-4}$ & $10^{-9}$ & $10^{-14}$ & $10^{-18}$ \\
Mathematica & $10^{1}$   & $10^{-1}$   & $10^{-6}$ & $10^{-10}$ & $10^{-15}$ & $10^{-19}$ \\
\noalign{\smallskip}\hline
\end{tabular}
\end{table}
\eex

If the data are given in units of 1, then $P=5$, $x_{\max}=33000$, $\bar c=0.33$ and $\bar P=5$ (Table \ref{t4}, second part). The polynomial function is then
\begin{equation} \label{E3}
 \begin{array}{rcl}
y^A_2&=&1.22036\cdot 10^{-14}x^5-7.8829\cdot 10^{-10}x^4+0.0181062\cdot 10^{-3} x^3\\
     &&- 0.32270\,x^2+ 10233.5\,x-8110000\,.
 \end{array}
 \end{equation}

 Corresponding coefficients of polynomials (\ref{E2}) and (\ref{E3}) have the same number of significant digits. They are just adjusted according to the chosen units of the experimental data. More precisely, let the experimental data $(x_i,y_i)$, $i=0,1,\dots,n$, are fitted by polynomial trend:
 \begin{itemize}
 \item $y^A(x)=\sum_{k=0}^{d} {\bf a_{k}}\,x^{k}$, where $x_i$ and $y_i$ are expressed in units of 1,
 \item $y^{u,A}(x)=\sum_{k=0}^{d} {\bf a^u_{k}}\,x^{k}$, where $x_i$ are expressed in units of $10^{u_x}$ and $y_i$ in units of $10^{u_y}$,
 \end{itemize}
 for $x\in [x_0,x_n]$. Then
 \begin{equation}\label{units}
 {\cal P}({\bf a_k})={\cal P}({\bf a^u_k})\cdot 10^{u_y-k\cdot u_x}\,,k=0,1,\dots,d\,.
 \end{equation}

 Note that (\ref{units}) holds for the coefficients obtained by Mathematica, too (Table \ref{t4}).

\subsection{Application of the algorithm on the fitting polynomial ($d<n$)}
Here, the proposed algorithm will be applied on the data from Hargreaves and McWilliams (\cite{harg10}) given in Table \ref{t5} (the first and second column).

\bex
The authors of \cite{harg10} reported 3rd order polynomial trend line (\ref{exc3rd}) with only one significant digit at the first term (obtained by Excel 2010).
\begin{equation}\label{exc3rd}
y^E_3(x)= - 3\cdot 10^{-5} x^3+ 0.0838\,x^2- 22.273\,x+9372.5
\end{equation}

\begin{table}[htbp]
\caption{Experimental data $(x_i,y_i)$ from \cite{harg10}, the absolute errors and coefficients of determination $R^2$ (E=Excel, S=Statistica,
T= from \cite{harg10}, M=Mathematica, A=Algorithm).}
\label{t5}       
{\scriptsize \begin{tabular}{rrrrrrr}
\hline\noalign{\smallskip}
$x_i$ & $y_i$ & $|y_i-y^E_3(x_i)|$& $|y_i-y^{S}_3(x_i)|$ & $|y_i-y^{T}_3(x_i)|$ & $|y_i-y^M_3(x_i)|$ & $|y_i-y^A_3(x_i)|$\\
\hline\noalign{\smallskip}
   0&10254&  881.5& 881.5& 881.5& 881.5& 882.0\\
 100&7577 &  376.2& 380.1& 380.2& 380.2& 379.7\\
 200&9723 & 1693.1&1661.8&1661.3&1661.3&1661.8\\
 300&5652 & 3770.6&3876.2&3877.4&3877.3&3876.9\\
 400&11310&  641.3& 891.6& 893.7& 893.7& 893.3\\
 500&19921& 4485.0&3996.1&3992.8&3992.8&3993.2\\
 600&17800& 1896.7&2741.6&2746.4&2746.2&2745.9\\
 700&25995& 1441.6&  99.9&  93.4&  93.6&  94.0\\
 800&36580& 6753.9&4751.1&4742.6&4742.9&4743.2\\
 900&34186& 1148.8&4000.5&4011.2&4010.9&4010.6\\
1000&44601& 3701.5& 210.3& 223.5& 223.1& 222.8\\
1100&49305& 2964.8&2241.8&2257.8&2257.3&2257.0\\
1200&62692&11215.1&4455.4&4436.4&4437.1&4437.4\\
1300&63541& 7411.4&1183.0&1205.2&1204.4&1204.2\\
1400&68413& 8294.7&2439.6&2465.3&2464.3&2464.1\\
1500&76650&13387.0& 184.3& 154.8& 156.0& 156.2\\
1600&84247&18863.3&2840.1&2806.5&2807.9&2808.0\\
1700&88477&22176.6&2957.3&2919.4&2921.1&2921.3\\
1800&84852&19018.9&3795.5&3838.0&3836.0&3835.8\\
1900&94884&31082.2&4250.2&4202.9&4205.3&4205.4\\
2000&89480&29453.5&1842.1&1894.5&1891.7&1891.6\\
2100&88259&33931.8&2296.8&2354.6&2351.3&2351.2\\
2200&86899&40375.1&1279.4&1342.8&1339.1&1339.0\\
2300&84281&47844.4& 247.6& 178.3& 182.6& 182.6\\
2400&76709&52823.7&1255.3&1330.7&1325.8&1325.8\\
2500&73854&65164.0&4039.5&3957.8&3963.2&3963.2\\
2600&58270&67599.3&1157.5&1245.9&1239.7&1239.7\\
\hline\noalign{\smallskip}
\multicolumn{2}{c}{$R^2(\%)$}&10.2172&99.2289&99.2302&99.2301&99.2303\\
\noalign{\smallskip}\hline
\end{tabular}
}
\end{table}

As the fit of the reported equation (\ref{exc3rd}) is not appropriate (Table \ref{t5}, third column, coefficient of determination is $10.22\%$, while the reported $R^2$ by Excel is $99.24\%$!), Hargreaves and McWilliams (\cite{harg10}) suggested that the polynomial coefficients should have five significant digits as given in (\ref{tom3rd}).
\begin{equation}\label{tom3rd}
y^T_3(x)= - 0.26088\cdot 10^{-4} x^3+ 0.083813\,x^2- 22.273\,x+9372.5
\end{equation}

This output of the polynomial is similar with the default output of the fitting polynomial obtained by Mathematica (\ref{math3rd}), where all the coefficients, except the constant term, have six significant digits.
\begin{equation}\label{math3rd}
y^M_3(x)=- 0.260884\cdot 10^{-4} x^3+ 0.0838132\,x^2- 22.2732\,x+ 9372.5\,.
\end{equation}

According to Table \ref{t5}, one can conclude that $P=0$. Since $y_{\max}=94884$ and $1-R^2=0.0077$, then $(1-R^2)\cdot y_{\max}=730.607$, so by definition \ref{def1} $(ii)$, $B=2$. From $x_{\max}=2600=0.26\cdot 10^{4}$ it follows $\bar c=0.26$, $\bar P=4$ and on the basis of the proposed algorithm, it is obtained the third order polynomial (\ref{alg3rd}),
\begin{equation}\label{alg3rd}
y^A_3(x)= - 0.26088\cdot 10^{-4} x^3 + 0.083813\,x^2 - 22.27\,x + 9370,
\end{equation}
where the precision of the coefficients ${\bf a_k}$, $k=0,1,2,3$ is equal to $10^{[P-1+B-k\,\bar P-P_{\bar c}]}$ (Table \ref{t6}).

\begin{table}[htbp]
\caption{The precision of the optimal coefficients of polynomial (\ref{alg3rd}), $P_{\bar c}=k\,\log_{10}\,\bar c$, $k=0,1,2,3$.}
\label{t6}       
\begin{tabular}{ccccc}
\hline\noalign{\smallskip}
{\scriptsize $P=0$, $B=2$, $\bar c=0.26$, $\bar P=4$} & ${\cal P}({\bf a_0})$ & ${\cal P}({\bf a_1})$ & ${\cal P}({\bf a_2})$ & ${\cal P}({\bf a_3})$\\
\noalign{\smallskip}\hline\noalign{\smallskip}
$10^{[P-1+B-k\,\bar P-P_{\bar c}]}$ & $10^{1}$   & $10^{-2}$   & $10^{-6}$ & $10^{-9}$  \\
\noalign{\smallskip}\hline
\end{tabular}
\end{table}

The default output of the fitting polynomial in Statistica 2012 reports that constant term 9372.4999 has even eight significant digits (on the basis of the proposed algorithm three is enough), which is completely unnecessary because the output data have order $10^4$ implying that all digits after decimal point in the constant term are redundant.
\eex

\subsection{The "First aid"}
We agree with Hargreaves and McWilliams (\cite{harg10}) that, until the developers of Excel do not fix the default output, the users should right-click on the fitted equation and select 'Format Trandline' \dots The users then has the option to display the equation using either standard decimal or scientific notation with as many displayed decimal places as desired. In Statistica, users should select 'Statistics$\backslash$Advanced Linear/Nonlinear Models$\backslash$Nonlinear Estimation...' and double-click on the graph.  On the other hand, transfer of values in Mathematica was very good and default output had high precision. Copy-Paste tool for the number transfer maintains the precision.

\section{Summary and conclusions}
We offer a simple recommendation. If it is possible, a selected software package should be used for all the research studies and then those studies should be checked by the software of different origin. This step allows the reduction of errors made by the users of software application, or the software application itself, to minimum. Eventually, there is no perfect software support.

Based on the above discussion, it can be concluded that:
 \begin{itemize}
 \item[1.] Presented default equations of polynomial functions do not correspond to the presented graphics of functions (Excel, Statistica).
 \item[2.] The default output of Mathematica shows more significant digits than necessary.
 \item[3.] Based on the precision of experimental data, proposed algorithm automatically determines the necessary and sufficient number of significant digits of the coefficient of polynomial trends. Thus, it is avoided the meaningless default outputs of the coefficients of polynomial trends. Also, the user does not have to manually check which number of significant digits gives the best data approximation.
 \item[4.] If the large accuracy of the output data is not necessary, it should be allowed that the user has the option to choose the precision of output data, i.e. to choose $P$, which may be less than the precision of the experimental data. For example, if the output value should be divided into several intervals, e.g. high, middle and low income, etc.
 \item[5.] If the data are fitted by the function other than polynomial, in general, applying the well known methods (logarithmic, square root, \dots), the function can be transform to the polynomial function, and then, after the usage of algorithm, the inverse transformation brings back the original form of the fitting function.
 \end{itemize}

 \section{Note}
 Violation of the order of operation as defined in mathematics was noticed during the analysis of experimental data and determination of thermal power of boiler plant for the combustion of wheat straw bales (\cite{ded08,ded10,ded12}). In the conducted research, the measurement results were automatically recorded in a controller every five seconds. The boiler thermal power was determined by using a direct method \cite{brk06}.

According to the experimental data, and for the air flow of 220 ${\rm m}^3{\rm h}^{-1}$, it was concluded that these data should be approximated by an exponential function  $y=d+c\,\exp^{-(b\,(x-a))^2}$ which would be symmetric with respect to the line $x=a$, and its graphic interpretation ("bell-shaped") fitted normal distribution ($a$ serves for graph translation along the $x$-axis). The coefficient $c$ influences the function maximum, $b$ on the bell width and $d$ on the translation along $y$-axis. Mathematical model of the correlation between boiler thermal power $P$(W) and bale incineration time $\tau$(s) is:
\begin{equation}\label{modelwheat}
P(\tau)=b_1+b_2\cdot e^{-\left(b_3\cdot(\tau-b_4)\right)^2}
\end{equation}
In this equation, $b_1$, $b_2$, $b_3$ and $b_4$ stand for the regression coefficients.

 After applying nonlinear regression analysis, the following regression coefficients of model (\ref{modelwheat}) were obtained for $95\%$ confidence level and $F$-value of $10242.3$ (Table \ref{tab1}).
\begin{table}[htbp]
\caption{Statistical analysis of the regression coefficients of (\ref{modelwheat}).}
\label{tab1}       
\begin{tabular}{ccccc}
\hline\noalign{\smallskip}
 & Estimate & Stand. err. & t-value & Confidence interval\\
\noalign{\smallskip}\hline\noalign{\smallskip}
$b_1$ & 43.5 & 0.8502 & 51.2 & (41.8, 45.2)\\
$b_2$ & 41.6 & 0.9484 & 43.4 & (39.3, 43.0)\\
$b_3$ & $37.2\cdot 10^{-4}$ & 0.0001 & 30.7 & $(35.4\cdot 10^{-4}, 40.0\cdot 10^{-4})$\\
$b_4$ & 443.8 & 3.6262 & 122.4 & (436.7, 451.0)\\
\noalign{\smallskip}\hline
\end{tabular}
\end{table}

\begin{figure}[htbp]
  \includegraphics[width=0.7\textwidth]{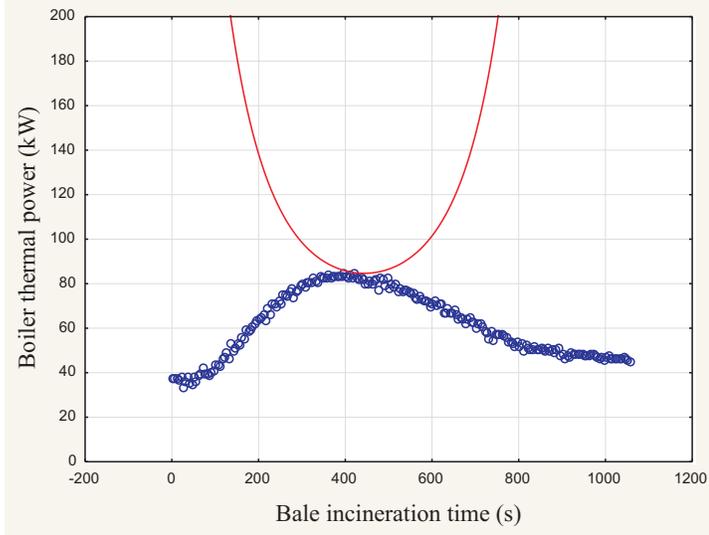}
\caption{Incorrectly drawn function (solid convex line) and experimental data (circles) for the air flow of 220 ${\rm m}^3{\rm h}^{-1}$ - Statistica.}
\label{incor}       
\end{figure}

Figure \ref{incor} shows the experimental data and fitting function. In this case, $R^2$ is $91\%$ and it can be concluded that the fitting function was incorrectly drawn by Statistica.
%

\begin{figure}[htbp]
  \includegraphics[width=0.6\textwidth]{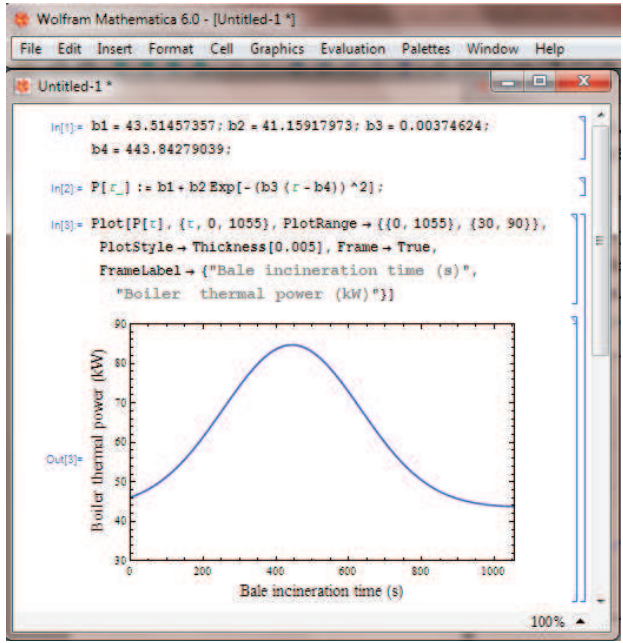}
\caption{Fitting function for the air flow of 220 ${\rm m}^3{\rm h}^{-1}$ - Mathematica.}
\label{p4}       
\end{figure}

The graph of the same function in Mathematica is given in Figure \ref{p4}. Since we were not able to find the reason for this mistake, we decided to fit the experimental data in Excel as a form of simpler software package. The fitting function was again incorrectly drawn.
%

\begin{figure}[htbp]
  \includegraphics[width=0.6\textwidth]{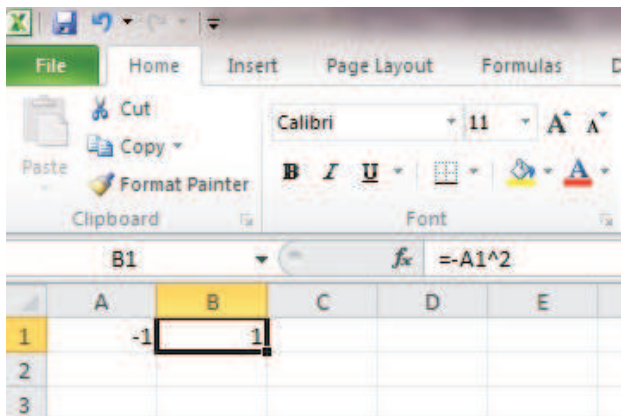}
\caption{Calculation of $-x^2$ by Excel. The order of operation is violated.}
\label{p1}       
\end{figure}

\begin{figure}[htbp]
  \includegraphics[width=0.7\textwidth]{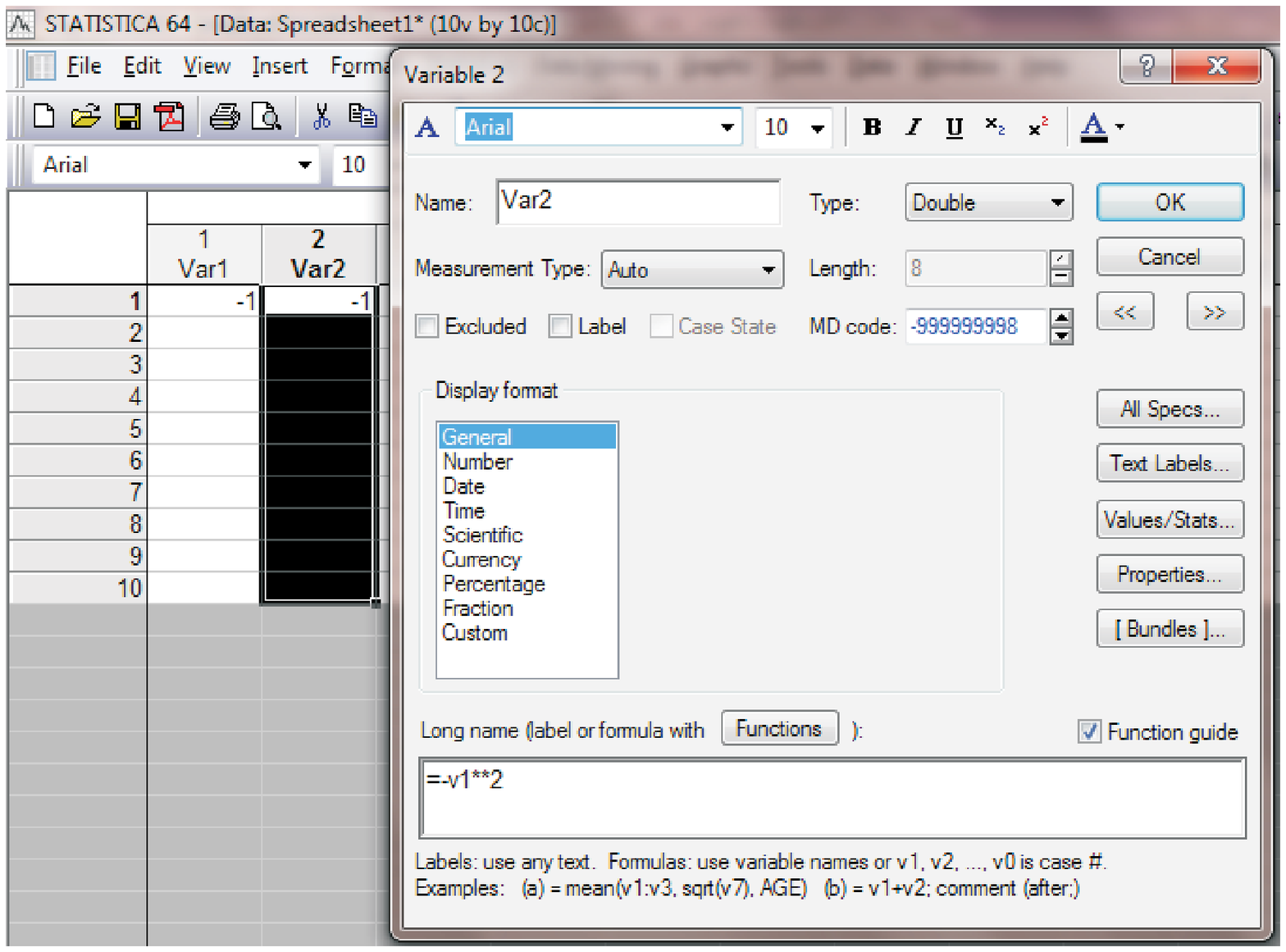}
\caption{Calculation of $-x^2$ by Statistica.}
\label{p3}       
\end{figure}

 The next step was to check the values of the fitting function at some points and that was the key moment. Program Excel violated the operation order. This arose from the fact that Excel calculated $-x^2$ as $(-x)^2$ as it can be seen from Figure \ref{p1}.
%
 Although Statistica correctly calculates $-x^2$ as $-(x^2)$ (Figure \ref{p3}), during the graphic presentation it again calculated $-x^2$ as $(-x)^2$ (Figure \ref{incor}).
%
The order of the operation has been violated once more. In comparison to model (\ref{modelwheat}), we had to add two new brackets in the exponent in order to obtain correct graphic presentation.

\begin{figure}[htbp]
  \includegraphics[width=0.6\textwidth]{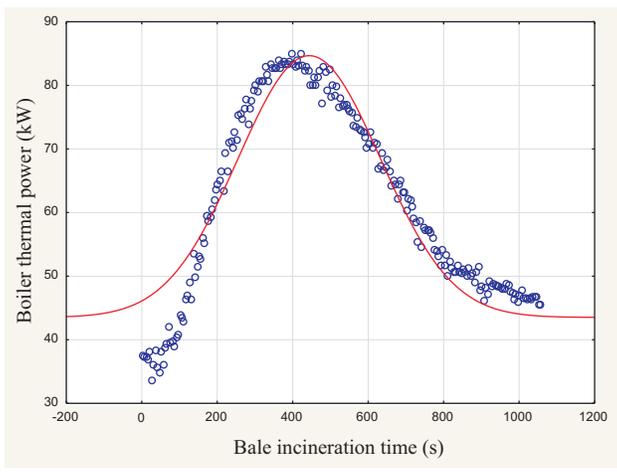}
\caption{Appropriate fitting function (solid curve) and experimental data (circles) for the air flow of 220 ${\rm m}^3{\rm h}^{-1}$ - Statistica.}
\label{2}       
\end{figure}

 By choosing the following model:
 $$
 P(\tau)=b_1+b_2\cdot e^{-\left(\left(b_3\cdot(\tau-b_4)\right)^2\right)}\,,
 $$
 the regression coefficients remain the same (Table \ref{tab1}), and the fitting function is now appropriate (Figure \ref{2}).
%
\newpage

Here, it should be said that all four packages correctly take $a-x^2$ as $a-(x^2)$, for $a\neq 0$, but calculation of $-x^{2k}$, $x\in\mR$, $k\in\mZ$ is still a problem in Excel and Statistica. For example, $-(-2)^4=16$, $-(-2)^{-4}=1/16$!\\

{\bf Acknowledgements}\\
This study was financed by the Ministry of Education, Science and Technological Development, the Republic of Serbia, Grant no. TR 37017 and TR 31046, 2011--2014.










\end{document}